\documentstyle[11pt,newpasp]{article}
 
\def\deg{$^{\circ}\,$}
\def\solm{M$_{\odot}\,$}

\def\etal{{\it et al.\ }}
\def\eg{{\it e.g.\ }}

\begin{document}

\title{Black Holes in Centers of Disk Galaxies --- \\ --- 
Spectroscopy with a Wide Slit}
\author{Witold Maciejewski} 
\affil{Theoretical Physics, University of Oxford}

\section{Introduction}
Motivated by STIS observations of more than 50 nearby galactic nuclei, 
we consider long-slit emission-line 
spectra when the slit is wider than the instrumental 
PSF. This practice is conventionally considered to enhance the 
signal-to-noise ratio (S/N) of the data at the price of what may be 
an insignificant loss in velocity resolution. However, when the target 
has arbitrarily large velocity gradients, the use of a wide slit can have 
more subtle effects, because the position and velocity information becomes 
entangled along the dispersion direction. Here we investigate these effects
for emission-line spectra from gaseous disks around a central massive 
black hole (BH) in galaxies, but they are applicable to any objects
with steep velocity gradients (\eg shocks or contact discontinuities).

\section{A slit wider than the PSF}
For a disk in circular motion in an axisymmetric galactic potential 
including a massive BH, a wide slit samples off-center velocities, 
as well as the central rotation curve (dashed line in Fig.1, left).
When the disk is observed in some emission line through a long-slit
spectrograph, the pattern of intensity resembles the left panel of Fig.1, 
but it is
modified by the geometry of spectrograph's optics: the diffraction pattern 
produced by light that enters near one edge of the slit is displaced with 
respect to light of the same frequency that enters at the corresponding 
point on the other edge of the slit. Thus the difference in position
across the slit is seen on the detector as an {\it instrumental velocity 
offset}. To reflect the light pattern on the detector, the left panel of Fig.1
has to be modified by shifting each sample line in the dispersion direction 
by a constant amount, different for each line. The result seen in 
the central panel of Fig.1 shows that in
the presence of a velocity gradient across the slit, the instrumental
velocity offset competes with the Doppler shifts: this 
offset wins at large radii, and at small radii the Doppler shifts take over.
In between, both factors are of similar strength: light 
from all positions across the slit converges at one effective 
velocity, forming {\it the caustic}. At radii interior to the caustic,
two maxima in the light distribution are present (Fig.1, right): one 
from the slit center (showing Keplerian rise), and one from the slit 
edges (passing through zero 
velocity at the nucleus). Thus we predict position-velocity diagrams for
rotating disks to be rich in structure; the information contained in 
it is lost if one merely fits a Gaussian emission line profile at each 
radius (which is the traditional approach).

\section{A new BH mass estimator}
The position of the caustic can indicate the presence, and betray the 
mass $M_\bullet$ of the BH. A wide slit contains many narrow slits 
within it, and it can provide information that otherwise would
need two off-set thin slits. If the caustic occurs at a position 
$\alpha$ down the $2\delta$-wide slit, then velocities at the slit 
center ($v_c$) and at the slit edge ($v_e$) are related by
$ v_c(\alpha) = v_e(\alpha) + B\delta $, where $B$ converts
the plate scale in the dispersion direction to the spectrograph's 
dispersion. This additional 
constraint allows us to recover the disk's inclination angle $i$ and $M_\bullet$
independently, rather than $M_\bullet \sin i$.
Moreover, this new method exploits an artifact at the outer edge of the 
BH's sphere of influence, and therefore gives higher sensitivity to BH 
detection than traditional methods based on the Keplerian rise in 
velocity occurring inside the sphere of influence.

\section{Confronting observations. Conclusions.}
The most detailed long-slit spectrum of nuclear emission so far
(M84, Bower \etal, 1998, {\it ApJ}, 492, L111) shows two light maxima 
at radii close to the nucleus. Although they were interpreted 
as coming from two physically distinct nuclear components, the
observed light pattern has the same structure as our models in Fig.1, and is
caused by a wide slit. We interpret the point where the track 
of maximum light splits into two as the caustic, from which we estimate 
the disk to be inclined at 74\deg, and the BH mass to be
$4 \times 10^8$ \solm, smaller than that of Bower \etal by a factor of 4.
BH masses derived by the traditional method may be overestimates (Maciejewski
\& Binney 2000, {\it MNRAS} submitted).

The finite width of the slit generates complex patterns in the spectra
that can have considerable diagnostic power if they are modeled with 
adequate sophistication. They allowed us to develop a new method for 
estimating the BH mass that gives fuller information and higher 
sensitivity to BH detection than traditional methods.

\begin{figure*}
\includegraphics{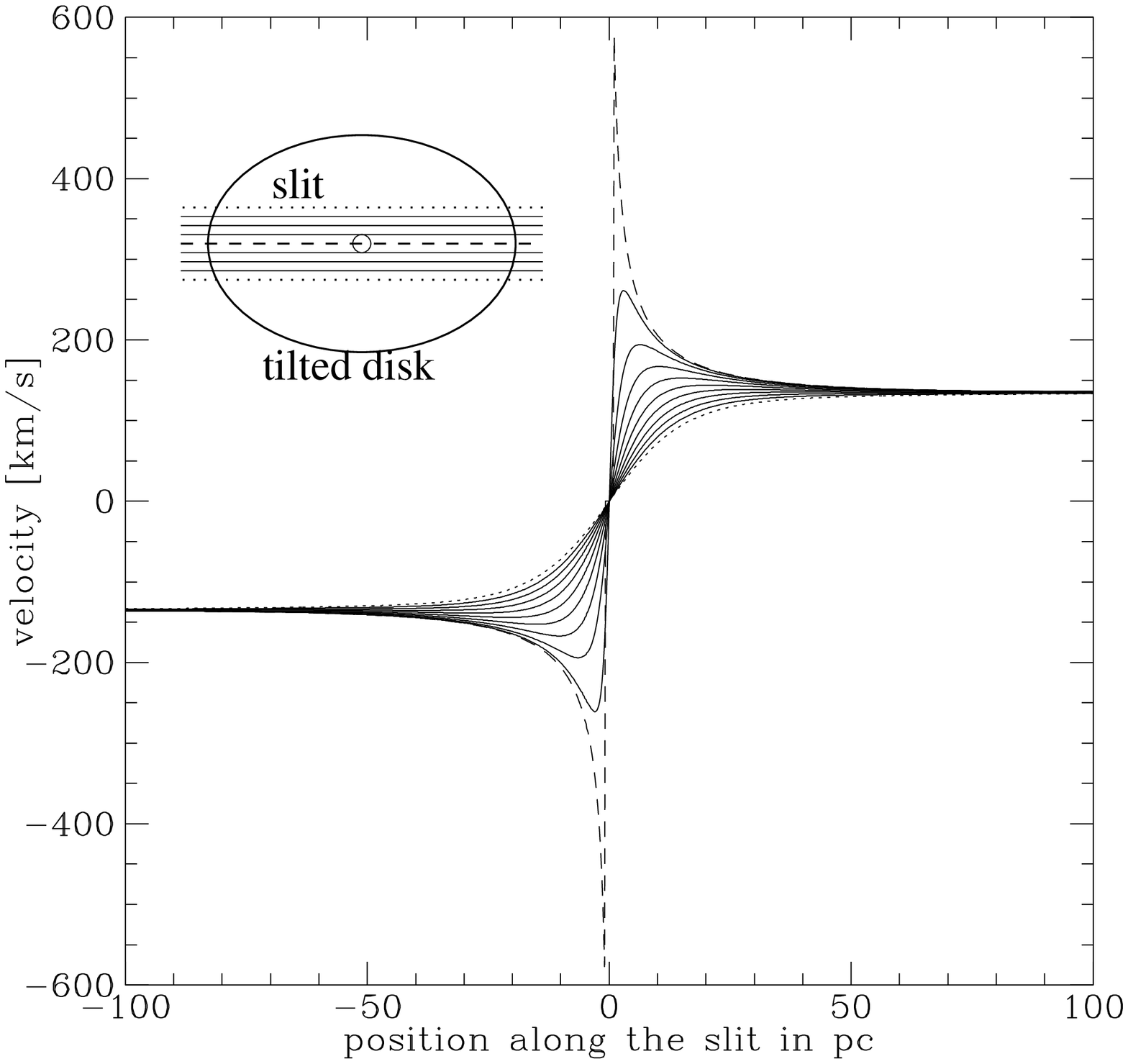}
\includegraphics{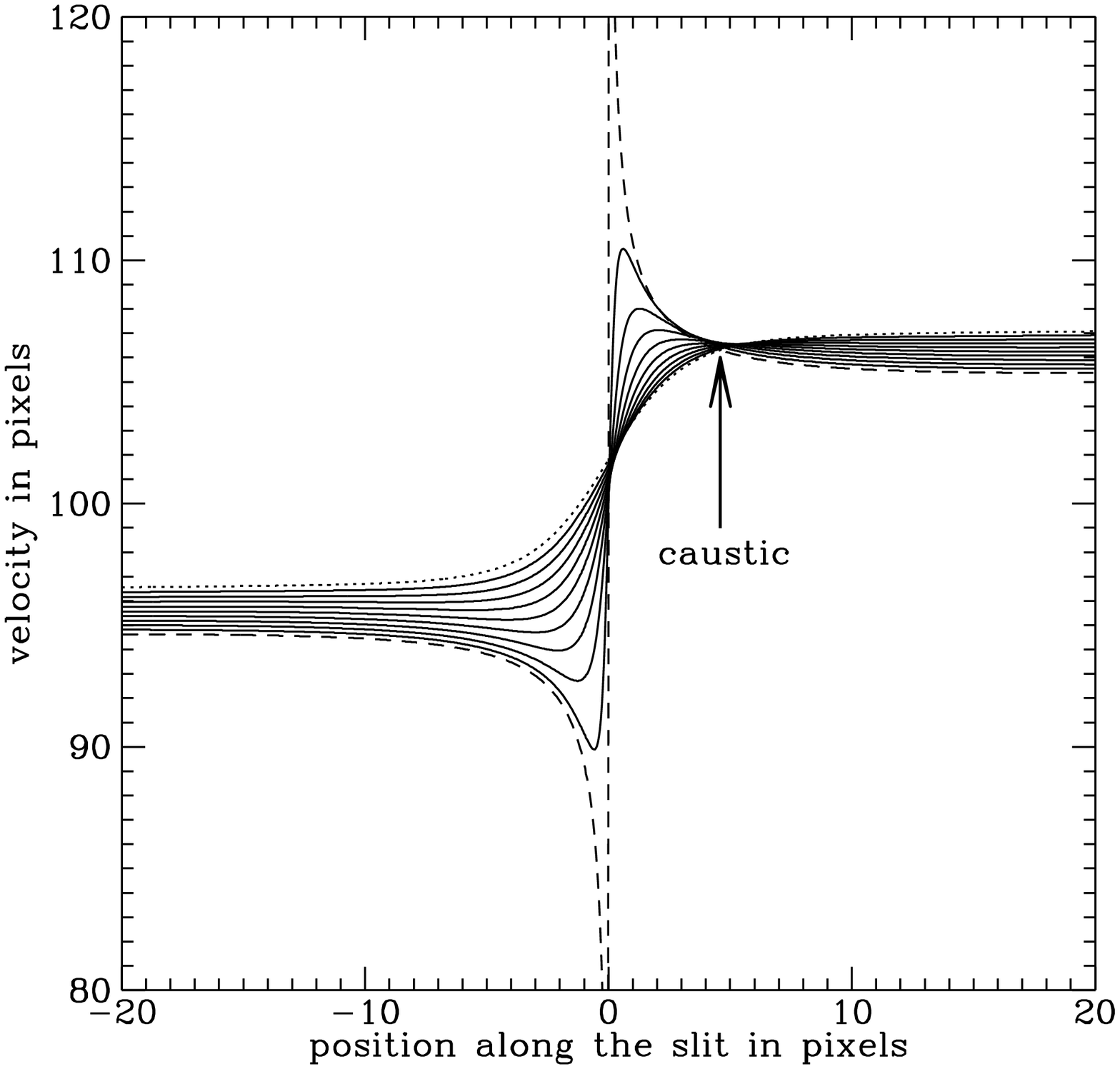}
\includegraphics{maciejewskiw3.ps}
\vspace{4cm}
\caption{{\it Left:} 
Position-velocity diagram for the disk inclined at 60\deg,
rotating in a potential of a $10^8$ \solm BH plus extended density
distribution $\sim R^{-1.8}$. Velocities are sampled along cuts parallel to 
the line of nodes, shown in the inset at upper left.
{\it Centre:} The light pattern on the spectrograph's detector
(only the top half of the slit is sampled)
{\it Right:} The light from entire slit integrated within the detector's pixels.}
\end{figure*} 

\end{document}